\documentclass[12pt]{article}
\usepackage{latexsym}
\usepackage{amsmath}
\usepackage{amssymb}
\usepackage{amsthm}
\usepackage{dsfont}
\usepackage{longtable}
\usepackage{color}
\usepackage{verbatim}
\usepackage[sort&compress,numbers]{natbib}
\usepackage{graphicx}
\usepackage{epstopdf}
\usepackage{float}
\usepackage{wrapfig}
\usepackage{ulem}

\newcommand {\dd}[2] {\frac {\partial {#1} }{\partial {#2}}}

\theoremstyle{definition}

\makeatletter
\@addtoreset{remark}{utv}
\@addtoreset{remark}{theorem}
\makeatother

\makeatletter
\renewcommand\@biblabel[1]{#1.}
\makeatother

\allowdisplaybreaks
\title{Multiparticle distributions and intercepts of $r$-particle correlation functions in the symmetric
Tamm-Dancoff type $q$-Bose gas model}
\author{A.M. Gavrilik and Yu.A. Mishchenko}

\date{\small\it Bogolyubov Institute for Theoretical Physics,
14-b, Metrolohichna str., Kiev, 03680, Ukraine\\
e-mail:}

\begin{document}
\maketitle

\begin{abstract}
Symmetric Tamm-Dancoff (STD) type $q$-deformed quantum oscillators
are used as the base for constructing the respective   %symmetric Tamm-Dancoff
 STD type $q$-Bose gas model. In this letter, within
the STD $q$-deformed Bose gas model we derive explicit analytic expressions
for the $r$-particle momentum distribution functions, and for the (momentum)
correlation function intercepts of 2nd, 3rd, and
any $r$th order. Besides, we obtain large-momentum asymptotic formulas for the $r$th
order intercepts which show dependence on the $q$-parameter only.
The obtained formulas provide new example,
in addition to already known two cases, of exact results for $r$-particle distributions
and  the correlation function intercepts in deformed analogs of Bose gas model.
 % as well within the deformed Bose gas model based on
 % symmetric Tamm-Dancoff deformation.
\end{abstract}

%{\bf Mathematics Subject Classification (2010).}
%17B37, %Nonassociative rings and algebras. Lie algebras and Lie superalgebras. Quantum groups (quantized enveloping algebras) and related deformations
%17B81, %Nonassociative rings and algebras. Lie algebras and Lie superalgebras. Applications to physics
%81R50, %Quantum theory. Groups and algebras in quantum theory. Quantum groups and related algebraic methods
%82B10, %Statistical mechanics, structure of matter. Quantum equilibrium statistical mechanics (general)
%82B21, %Statistical mechanics, structure of matter. Equilibrium statistical mechanics. Continuum models (systems of particles, etc.)
%82D05. %Statistical mechanics, structure of matter. Applications to specific types of physical systems. Gases

 \newline    %{\bf PACS}: 05.30.Jp; 05.90.+m; 11.10.Lm; 25.75.Gz; 02.30.Gp.

\section{Introduction}

The statistical (two-dimensional, lattice) models constitute large
class of exactly solvable models~\cite{Baxter} whose development is
witnessed even in present days. The solvability in those models
implies the possibility of finding exact results for the partition
function, correlation functions etc. On the other hand, there exists
another important branch of modern statistical physics, namely the
study of deformed analogs of Bose (and Fermi) gas model in two,
three, etc. dimensions. Its development, from the earliest papers on
the
subject~\cite{LeeYu1990,GeSu1991,Martin-Delgado1991,Chaichian1993,Man'ko1993}
and till some most recent
ones~\cite{AlginIlik2013,GM_UJP2013,Rovenchak2014} demonstrates that
this still remains a hot topic (see
also~\cite{DaiXie_rev1,AlginSenay2012} for some, by no means
exhaustive, list of works published during the last two decades),
with many interesting problems
awaiting for their resolving. Among these, it is worth to point out  %concentrate
the task of deriving exact expressions for statistical quantities,  %for thermodynamical
in particular the multi-particle distribution functions and
respective (momentum) correlation functions or their intercepts. At
present, to the best of our knowledge, for only two essentially
differing deformed analogs of the Bose gas model  that task has been
performed. Namely, the $r$-particle ($r\ge 1$) distribution
functions and the respective $r$-particle correlation function
intercepts have been obtained (i) for the $p,q$-deformed Bose gas
model, see~\cite{AdGa}, and (ii) for the recently proposed
$\mu$-deformed analog of Bose gas model~\cite{GaMi2012}. The two
models principally differ: while the first one belongs to Fibonacci
class~\cite{Arik1992Fib}, the second one is not Fibonacci, but is a
typical representative of the class of quasi-Fibonacci
models~\cite{GKR_Fib}.

Recently, a new symmetric Tamm-Dancoff (STD) $q$-deformed oscillator has
been introduced~\cite{Chung_S-TD} and some of its major properties studied.
Note that usual Tamm-Dancoff $q$-oscillator~\cite{OdakaTD1,Chaturvedi1993_TD2} involves only
{\it real values} of the parameter $q$, and possesses diverse cases of accidental
degeneracies of energy levels~\cite{GR_TD3}. The STD $q$-oscillator admits, besides
real, also the {\it complex phase-like} values of deformation parameter $q$ and, moreover, there
exist definite values of $q=\exp(\rm i\theta)$ for which accidental degeneracies
of energy levels do occur~\cite{Chung_S-TD}. The possibility to deal with the complex-valued
deformation parameter, which we have in particular in the STD type $q$-deformed
model, gives some important advantages, as it was discussed in (the last paragraph of) ref.~\cite{Chung_S-TD}.

Let us emphasize that whereas usual Tamm-Dancoff $q$-oscillator belongs to the class of
Fibonacci oscillators, the STD $q$-deformed oscillator
does not. Namely, as shown in~\cite{Chung_S-TD},  it is a quasi-Fibonacci one (the notion
of quasi-Fibonacci oscillators was introduced in~\cite{GKR_Fib}).

In this letter our goal is to derive, for the STD type $q$-deformed
Bose gas model based on the set of    %symmetric Tamm-Dancoff $q$-deformed
STD type $q$-deformed oscillators,  %% symmetric Tamm-Dancoff $q$-deformed oscillator
the exact expressions for the $r$-particle ($r\ge 1$) distribution
functions and the respective $r$-th order correlation function intercepts. By deriving
this result we demonstrate that there is one more, the third, $q$-deformed family of modified Bose-gas
like models for which it proves possible to obtain the exact analytical formulas
for the $r$th order distribution functions and $r$-particle correlation intercepts.

\section{Basics of the symmetric Tamm-Dancoff $q$-oscillator~\cite{Chung_S-TD}}\label{sec:std-basics}

The symmetric $q$-deformed bosonic Tamm-Dancoff oscillator algebra is defined as
\begin{align}
&aa^{\dagger} - a^{\dagger}a = \{N+1\}_q - \{N\}_q = \frac{1}{2} ( 1
+ (1-q^{-1} )N ) )
q^{N} +  \frac{1}{2} ( 1 + (1-q) N) q^{-N}, \label{1.1}\\
&[N, a^{\dagger}] = a^{\dagger}, \quad [N, a]= -a,\nonumber
\end{align}
where
\begin{equation} \label{phi_STD}
a^\dag a = \{N\}_q \equiv\varphi_{\rm STD}(N)\equiv \frac{N}{2}(q^{N-1}+q^{-N+1})= \frac{N}{2} q^{-N+1} (1 + (q^2)^{N-1} )
\end{equation}
with $\{N\}_q$ denoting the STD type $q$-number ($q$-bracket) or
structure function, and $q$ is either real, \ $0< q \le \infty$, or
complex phase-like: $q=\exp({\rm i}\theta)$, \  $-\pi \le\theta\le
\pi$.

One can easily show that the $q$-analog Fock-type representation of
the algebra (\ref{1.1}) is valid:
\begin{align*}
&N|n\rangle = n |n\rangle, ~~~~ n=0, 1, 2, \ldots ,\\
&a|n\rangle = \sqrt{\frac{ n }{2} ( q^{n-1} + q^{-n+1} )  }|n-1\rangle  = \sqrt{\{ n \}_q}|n-1\rangle,\\
%\label{1.3}
&a^\dag |n\rangle = \sqrt{\frac{ n+1 }{2} ( q^{ n} + q^{-n} )  }|n+1\rangle = \sqrt{\{n+1\}_q}|n+1\rangle.
\end{align*}
Remark that this STD $q$-bracket can be also written as a
``multiplicative hybrid'' of the structure function
$\varphi(n)\!=\!n$ of usual oscillator and that of the Biedenharn -
Macfarlane $q$-deformed oscillator~\cite{Bied,Macfar} i.e. the
(obviously symmetric under $q\leftrightarrow q^{-1}$) structure
function $\varphi_{BM}(n)\equiv[n]_q=\frac{q^n-q^{-n}}{q-q^{-1}}$,
as follows:
\begin{equation}
\{ n\}_q=n\,\frac{[n]_q\!-\![n\!-\!2]_q}{2} \qquad {\rm or} \qquad
 \{ n\}_q=n\,\frac{[2(n-1)]_q}{2\,[n-1]_q}.
\end{equation}

\section{Intercepts of 2nd and 3rd order correlation functions}\label{sec2}

The deformed Bose gas model constructed from the set of independent
modes of deformed oscillators with STD type structure function of
deformation $\varphi_{\rm STD}(N)$, see eq.~(\ref{phi_STD}), is
studied here. We start with the following defining expression for
the intercept (see e.g.~\cite{ChapmanHeinz1994} for a nondeformed
case, and~\cite{AdGa} for deformed one) of $r$th order momentum
correlation function, with fixed momentum~$\bf k$:
\begin{equation}\label{lambda_def}
\lambda^{(r)}({\bf k}) = \frac{\langle (a^\dag_{\bf k})^r (a_{\bf k})^r \rangle}{\langle a^\dag_{\bf k} a_{\bf k}\rangle^r}-1
= \frac{\langle \varphi(N_{\bf k})\varphi(N_{\bf k}-1)\cdot...\cdot\varphi(N_{\bf k}-r+1) \rangle}{\langle \varphi(N_{\bf k})\rangle^r}-1.
\end{equation}
The bracket $\langle...\rangle$ denotes statistical (thermal)
average.  As seen, to find the intercepts $\lambda^{(2)}({\bf k})$
and $\lambda^{(3)}({\bf k})$ we have to calculate the averages
$\langle a^\dag_{\bf k} a_{\bf k}\rangle$, $\langle (a^\dag_{\bf
k})^2 (a_{\bf k})^2 \rangle$ and $\langle (a^\dag_{\bf k})^3 (a_{\bf
k})^3 \rangle$. In what follows, when performing the calculations
that involve the quantities for a fixed mode, the index $\bf k$ will
be omitted for the sake of simplicity.

Taking the Hamiltonian in the simplest linear (additive) form
\begin{equation}\label{H}
H= \sum_{\bf k} \varepsilon({\bf k}) N_{\bf k} = \sum_{\bf k} \hbar\omega_{\bf k} N_{\bf k},
\end{equation}
for the average $\langle a^\dag a\rangle$ we find:
\begin{align}
\langle a^\dag a\rangle &= \langle\varphi(N)\rangle = \sum_{n=0}^\infty \frac{n}{2}(q^{n-1}+q^{-n+1}) e^{-\beta \hbar\omega n} \Bigl/
\sum_{n=0}^\infty e^{-\beta \hbar\omega n} = \nonumber\\
&= \frac12 (1-e^{-x}) \Bigl(-\dd{}{x}\Bigr) \sum_{n=0}^\infty (q^{n-1}+q^{-n+1}) e^{-nx}
= \frac{(1-e^{-x})}{2} \Bigl(-\dd{}{x}\Bigr) \Bigl(\frac{q^{-1}}{1\!-\!q e^{-x}} + \frac{q}{1\!-\!q^{-1} e^{-x}}\Bigr) =\nonumber\\
&= \frac{e^{-x}(1\!-\!e^{-x})}{2} \Bigl(\frac{1}{(1\!-\!q e^{-x})^2} + \frac{1}{(1\!-\!q^{-1} e^{-x})^2}\Bigr) \label{<a^dag_a>}
\end{align}
where $x = \beta \hbar\omega_{\bf k}$, $\beta=\frac{1}{k_B T}$,
$k_B$ is Boltzmann's constant. In a similar way we calculate
$\langle (a^\dag)^2 (a)^2 \rangle$:
\begin{align*}
\langle (a^\dag)^2 a^2 \rangle &= \langle\varphi(N)\varphi(N-1)\rangle = \frac14 \langle N(N-1)(q^{N-1}+q^{-N+1})(q^{N-2}+q^{-N+2})\rangle =\\
&= \frac14 (1-e^{-x}) \Bigl(\dd{^2}{x^2}+\dd{}{x}\Bigr) \sum_{n=0}^\infty (q^{n-1}+q^{-n+1})(q^{n-2}+q^{-n+2}) e^{-nx} =\\
&= \frac14 (1-e^{-x}) \Bigl(\dd{^2}{x^2}+\dd{}{x}\Bigr) \sum_{n=0}^\infty (q^{2n-3} + q^{-2n+3}+q^{-1}+q) e^{-nx}=\\
&= \frac14 (1-e^{-x}) \Bigl(\dd{^2}{x^2}+\dd{}{x}\Bigr) \Bigl(\frac{q^{-3}}{1-q^2e^{-x}} + \frac{q^3}{1-q^{-2}e^{-x}} + \frac{q+q^{-1}}{1-e^{-x}}\Bigr).
\end{align*}
Utilizing the equality $\Bigl(\dd{^2}{x^2}+\dd{}{x}\Bigr)
\frac{1}{1-\alpha e^{-x}}  = 2\frac{\alpha^2 e^{-2x}}{(1-\alpha e^{-x})^3}$ we
arrive at the desired expressions
\begin{equation}\label{r=2}
\begin{aligned}
&\langle (a^\dag)^2 a^2 \rangle = \frac12 e^{-2x} (1-e^{-x}) \Bigl[\frac{q}{(1-q^2e^{-x})^3} + \frac{q^{-1}}{(1-q^{-2}e^{-x})^3}
+ \frac{q+q^{-1}}{(1-e^{-x})^3}\Bigr],\\
&\lambda^{(2)}({\bf k}) = \frac{2 \bigl[\frac{q}{(1-q^2e^{-x})^3} + \frac{q^{-1}}{(1-q^{-2}e^{-x})^3}
+ \frac{q+q^{-1}}{(1-e^{-x})^3}\bigr]}{(1\!-\!e^{-x}) \bigl((1\!-\!q e^{-x})^{-2} + (1\!-\!q^{-1} e^{-x})^{-2}\bigr)^2}-1.
\end{aligned}
\end{equation}
Analogously, with account of equality
$\Bigl(\dd{}{x}+2\Bigr)\Bigl(\dd{}{x}+1\Bigr)\dd{}{x}\,
\frac{1}{1-\alpha e^{-x}} = -6\frac{\alpha^3 e^{-3x}}{(1-\alpha
e^{-x})^4}$, for $\langle (a^\dag)^3 a^3 \rangle$ we obtain the
third order distribution function and same order correlation
function intercept:
\begin{equation}\label{r=3}
\begin{aligned}
&\langle (a^\dag)^3 a^3 \rangle = \frac34 e^{-3x} (1\!-\!e^{-x}) \Bigl[\frac{q^3}{(1\!-\!q^3e^{-x})^4} + \frac{q^{-3}}{(1\!-\!q^{-3}e^{-x})^4}
+ \frac{q^3(1\!+\!q^{-2}\!+\!q^{-4})}{(1\!-\!qe^{-x})^4} + \frac{q^{-3}(1\!+\!q^2\!+\!q^4)}{(1\!-\!q^{-1}e^{-x})^4}\Bigr],\\
&\lambda^{(3)}({\bf k}) = \frac{6 \bigl[\frac{q^3}{(1\!-\!q^3e^{-x})^4} + \frac{q^{-3}}{(1\!-\!q^{-3}e^{-x})^4}
+ \frac{q^3(1\!+\!q^{-2}\!+\!q^{-4})}{(1\!-\!qe^{-x})^4} + \frac{q^{-3}(1\!+\!q^2\!+\!q^4)}{(1\!-\!q^{-1}e^{-x})^4}\bigr]}{(1\!-\!e^{-x})^2 \bigl((1\!-\!q e^{-x})^{-2} + (1\!-\!q^{-1} e^{-x})^{-2}\bigr)^3}-1.
\end{aligned}
\end{equation}

\section{Distributions and correlation function intercepts of $r$th order}\label{sec3}

The definition~(\ref{lambda_def}) of $r$th order intercepts
$\lambda^{(r)}$ involves the averages (deformed distributions)
$\langle (a^\dag)^r a^r \rangle$, that in the STD $q$-Bose gas case
gives
\begin{equation}
\langle (a^\dag)^r a^r \rangle = \langle \varphi(N)\cdot...\cdot\varphi(N-r+1) \rangle
= \langle N(N-1)...(N-r+1) \prod_{j=1}^r (q^{N-j}+q^{-N+j})\rangle.
\end{equation}
For the product appearing in this equality we obtain
\begin{multline} \label{prod_q^n}
\prod_{j=1}^r (q^{N-j}+q^{-N+j}) = \prod_{j=1}^r q^{N-j} \prod_{j=1}^r (1+(q^2)^{-N+j})
= q^{r N - r(r+1)/2} \sum_{k=0}^r (q^2)^{-kN} \sum_{1\le j_1<...<j_k\le r} (q^2)^{j_1+...+j_k} =\\
 = q^{r N - r(r+1)/2} \sum_{k=0}^r (q^2)^{-kN}\sum_{s=k(k+1)/2}^{(2r-k+1)k/2} p(k,r,s) q^{2s}
\end{multline}
where $p(k,r,s)$ is the number of partitions of $s$ into $k$ distinct summands each of
which being not greater than $r$. Remark also that the product in~(\ref{prod_q^n}) can be
expressed through the Gaussian $q$-binomial coefficients $\Bigl({\scriptstyle n\atop \scriptstyle k}\Bigr)_q$
(see e.g.~\cite{KacCheung} for definition) by means of the identity
\begin{equation}
\prod_{j=1}^r (1+(q^2)^{-N+j}) = \prod_{j=0}^{r-1} (1+q^{-2N+2}\cdot q^{2j}) = \sum_{k=0}^r q^{k(k+1)}
\Bigl({\scriptstyle r\atop \scriptstyle k}\Bigr)_{q^2} (q^2)^{-kN}.
\end{equation}
Now perform the calculation that generalizes the case of $\langle (a^\dag)^2 a^2 \rangle$
or $\langle (a^\dag)^3 a^3 \rangle$ to any $r$th order:
\begin{multline}\label{eq1}
\langle (a^\dag)^r a^r \rangle = \langle N(N-1)...(N-r+1)\prod_{j=1}^r (q^{N-j}+q^{-N+j})\rangle
= (1-e^{-x}) \sum_{n=0}^\infty \frac{1}{2^r} n(n-1)...(n-r+1)\cdot\\
\cdot\prod_{j=1}^r (q^{n-j}+q^{-n+j}) e^{-nx}
= \frac{(-1)^r}{2^r} (1-e^{-x}) \Bigl(\dd{}{x}+r-1\Bigr)...\dd{}{x} \sum_{n=0}^\infty \prod_{j=1}^r
(q^{n-j}+q^{-n+j}) e^{-nx} =\\
= \frac{(-1)^r}{2^r} (1-e^{-x}) \Bigl(\dd{}{x}+r-1\Bigr)...\dd{}{x}
\sum_{n=0}^\infty q^{r n - r(r+1)/2} \sum_{k=0}^r (q^2)^{-kn} \sum_{s=k(k+1)/2}^{(2r-k+1)k/2} p(k,r,s) q^{2s} e^{-nx} =\\
= \frac{(-1)^r}{2^r} (1-e^{-x}) \sum_{k=0}^r \sum_{s=k(k+1)/2}^{(2r-k+1)k/2} p(k,r,s) q^{2s-r(r+1)/2} \Bigl(\dd{}{x}+r-1\Bigr)...\dd{}{x} \frac{1}{1-q^{r-2k} e^{-x}}.
\end{multline}
By induction, it can be verified that
\begin{equation}\label{n-der}
\Bigl(\dd{}{x}+r-1\Bigr)...\Bigl(\dd{}{x}+1\Bigr)\dd{}{x}\,
\frac{1}{1-q e^{-x}} = (-1)^r r! \frac{q^r e^{-rx}}{(1-q
e^{-x})^{r+1}}.
\end{equation}
Indeed,
\begin{multline}
\Bigl(\dd{}{x}+r\Bigr)...\dd{}{x}\, \frac{1}{1-q e^{-x}} = (-1)^r r! q^r \Bigl(\frac{-r e^{-rx}}{(1-q e^{-x})^{r+1}} - (r+1) \frac{q e^{-x} e^{-rx}}{(1-q e^{-x})^{r+2}}\Bigr) + (-1)^r r! r \frac{q^r e^{-rx}}{(1-q e^{-x})^{r+1}}
=\\
= (-1)^{r+1} (r+1)! \frac{q^{r+1} e^{-(r+1)x}}{(1-q e^{-x})^{r+2}}.
\end{multline}
From (\ref{n-der}) and (\ref{eq1}) we derive one of our major formulas, namely
\begin{multline}
\langle (a^\dag)^r a^r \rangle = \frac{r!}{2^r} (1-e^{-x})e^{-rx} \sum_{k=0}^r \sum_{s=k(k+1)/2}^{(2r-k+1)k/2} p(k,r,s) q^{2s-r(r+1)/2} \frac{q^{(r-2k)r}}{(1-q^{r-2k} e^{-x})^{r+1}} =\\
= \frac{r!}{2^r} (1-e^{-x})e^{-rx} \sum_{k=0}^r \frac{q^{(r-2k)r}}{(1-q^{r-2k} e^{-x})^{r+1}} \Bigl(\sum_{s=k(k+1)/2}^{(2r-k+1)k/2} p(k,r,s) q^{2s-r(r+1)/2}\Bigr)
\end{multline}
which in terms of well-known $q$-binomials takes the form
\begin{equation}\label{<(a^+)^r_a^r>}
\langle (a^\dag)^r a^r \rangle =  \frac{r!}{2^r} (1-e^{-x})e^{-rx}
\sum_{k=0}^r q^{k(k+1)-r(r+1)/2} \Bigl({\scriptstyle r\atop
\scriptstyle k}\Bigr)_{q^2} \frac{q^{(r-2k)r}}{(1-q^{r-2k}
e^{-x})^{r+1}}.
\end{equation}
Using~(\ref{<a^dag_a>}), (\ref{<(a^+)^r_a^r>}) and definition~(\ref{lambda_def}) we
write out the expression for the $r${\it th} order intercept as our final result:
\begin{equation}\label{lambda_r}
\lambda^{(r)}({\bf k}) = \frac{r! q^{-r(r-1)/2}\sum_{k=0}^r
\bigl({\scriptstyle r\atop \scriptstyle k}\bigr)_{q^2}
\frac{q^{(k-r)(k-r+1)}}{(1-q^{r-2k}
e^{-x})^{r+1}}}{(1\!-\!e^{-x})^{r-1} \bigl((1\!-\!q e^{-x})^{-2} +
(1\!-\!q^{-1} e^{-x})^{-2}\bigr)^r} - 1 , \hspace{12mm}
 x = \beta \hbar\omega_{\bf k}.
\end{equation}
Its $\beta\hbar\omega_{\bf k}\to\infty$ (large momentum or low
temperature) asymptotics takes the form:
\begin{equation}\label{lambda_r_as}
\lambda^{(r)}_{as} = \{r\}_q!-1 =  \frac{r!}{2^r} \prod_{k=1}^r (q^{k-1}+q^{-k+1}) - 1.
\end{equation}
Let us stress that the obtained asymptotics depends, besides the order $r$, on the deformation
parameter $q$ {\it only}: neither mass of particle nor the temperature of $q$-Bose gas survive
in the asymptotics. In the particular case of $r=4$ we have:
\begin{multline}\label{r=4}
\langle (a^\dag)^4 a^4 \rangle = \frac32 (1-e^{-x})e^{-4x} \Bigl[\frac{q^6}{(1-q^4e^{-x})^5} + \frac{1+q^2+q^4+q^6}{(1-q^2e^{-x})^5} + \frac{q^{-4}+q^{-2}+2+q^2+q^4}{(1-e^{-x})^5} + \\
+ \frac{1+q^{-2}+q^{-4}+q^{-6}}{(1-q^{-2}e^{-x})^5} + \frac{q^{-6}}{(1-q^{-4}e^{-x})^5}\Bigr],
\end{multline}
and
\begin{equation}
\lambda^{(4)}({\bf k}) = \frac{24 \bigl[\frac{q^6}{(1-q^4e^{-x})^5} + \frac{1+q^2+q^4+q^6}{(1-q^2e^{-x})^5} + \frac{q^{-4}+q^{-2}+2+q^2+q^4}{(1-e^{-x})^5} + \frac{1+q^{-2}+q^{-4}+q^{-6}}{(1-q^{-2}e^{-x})^5} + \frac{q^{-6}}{(1-q^{-4}e^{-x})^5}\bigr]}{(1\!-\!e^{-x})^3 \bigl((1\!-\!q e^{-x})^{-2} + (1\!-\!q^{-1} e^{-x})^{-2}\bigr)^4}-1.
\end{equation}

\vspace{12mm}
It is worth noting that the symmetry under $q\to q^{-1}$, though
certainly valid, is not so obvious in the obtained general result
(\ref{<(a^+)^r_a^r>}), but, it is easily seen for each of the particular results
(\ref{r=2}), (\ref{r=3}), (\ref{r=4}) for $r=2, 3, 4$ respectively.

As a kind of consistency check, we take the $q\to 1$ limit of
general formula (\ref{<(a^+)^r_a^r>}) and obtain
\begin{equation}
\langle (a^\dag)^r a^r \rangle |_{q\to 1}=
 \frac{r!}{2^r} %(1-e^{-x})
 \, e^{-rx} \sum_{k=0}^r \Bigl({\scriptstyle r\atop
\scriptstyle k}\Bigr) \frac{1}{(1- e^{-x})^{r}} =
 \frac{{r!}\, e^{-rx}}{(1-e^{-x})^r},\qquad \lambda^{(r)}_{as}|_{q\to 1} = r!-1,
\end{equation}
as it should be in the usual case of standard (ideal) Bose gas model.

Thus, the expressions (\ref{<(a^+)^r_a^r>})-(\ref{lambda_r_as})
constitute our {\it exact results} for the $r$-particle momentum
distributions and $r$-th order ($r\ge 2$) correlation function
intercepts, along with their large momentum asymptotics, established
in the symmetric TD type $q$-Bose gas model.

\vspace{12mm}

\section{Concluding remarks}

The results obtained in this letter (presented in Eqs.~(\ref{<a^dag_a>}), (\ref{<(a^+)^r_a^r>}), (\ref{lambda_r}), (\ref{lambda_r_as}))
represent the third particular case from among different deformed analogs of Bose gas model
wherein the exact expressions for the $r$-particle distribution
functions and for the respective $r$th order correlation functions
intercepts have been derived. It is of interest to compare (the form
of) these formulas with the two analogous previous results for the
$p,q$-Bose gas model and for the $\mu$-deformed analog of Bose gas
model. Whereas in ref.~\cite{AdGa} the $r$th order correlation intercept $\lambda^{(r)}_{p,q}+1$
appears as fully factorized one-term expression depending on $e^x\equiv e^{\beta\hbar\omega}$
and $p,\,q$, the analogous results for $\lambda^{(r)}_\mu+1$ in~\cite{GaMi2012} and $\lambda^{(r)}_q+1$
in the present work are obtained as non-factorized expressions consisting of $r+1$ terms, each
of which formed from elementary functions of $e^{\beta\hbar\omega}$ in this letter (see eq.~(\ref{lambda_r})
above), but, each one containing special function (Lerch transcendent) in ref.~\cite{GaMi2012}.
Anyway, with these particular three models of deformed Bose gas at hands,
we have to mention in conclusion that now an interesting problem arises of describing
the whole class of deformed Bose gas models which admit the obtaining of similar exact results.

\section*{Acknowledgements}
This work was partly supported by the Special Program of the Division of
Physics and Astronomy of NAS of Ukraine and by the Grant (Yu.A.M.)
for Young Scientists of NAS of Ukraine (No.~0113U004910).

%\nocite{*}
%\bibliographystyle{apsrev4-1}


\begin{thebibliography}{10} %{references}
 \bibitem{Baxter} Baxter, R. J.: Exactly Solved Models in Statistical Mechanics, Dover Publications, Mineola, N.Y. (2007).
 \bibitem{LeeYu1990} Lee, C.R., Yu, J.-P.: On $q$-analogues of the statistical distribution, Phys. Lett. A {\bf 150}, 63-66 (1990).
 \bibitem{GeSu1991} Ge, M.-L., Su, G.: The statistical distribution function of the $q$-deformed harmonic oscillator, J. Phys. A: Math. Gen. {\bf 24}, L721-L723 (1991).
 \bibitem{Martin-Delgado1991} Martin-Delgado, M. A.: Planck distribution for a $q$-boson gas, J. Phys. A: Math. Gen. {\bf 24}, L1285-L1291 (1991).
 \bibitem{Chaichian1993} Chaichian, M., Felipe, R.G., Montonen, C.: Statistics of $q$-oscillators, quons and relations to fractional statistics, J. Phys. A: Math. Gen. {\bf 26}, 4017-4034 (1993).
 \bibitem{Man'ko1993} Man'ko V. I. et al: Correlation functions of quantum $q$-oscillators, Phys. Lett. A {\bf 176}, 173-175 (1993).
 \bibitem{AlginIlik2013} Algin, A., Ilik, E.: Low-temperature thermostatistics of Tamm-Dancoff deformed boson oscillators, Phys. Lett. A {\bf 377}, 1797-1803 (2013).
 \bibitem{GM_UJP2013} Gavrilik, A.M., Mishchenko, Yu.A.: Deformed Bose gas models aimed at taking into account both compositeness of particles and their interaction, Ukr. J. Phys. {\bf 58}, 1171-1177 (2013).
 \bibitem{Rovenchak2014} Rovenchak, A.: Complex-valued fractional statistics for $D$-dimensional harmonic oscillators, Phys. Lett. A {\bf 378}, 100-108 (2014).
 \bibitem{DaiXie_rev1} Dai, W.-S., Xie, M.: Calculating statistical distributions from operator relations: The statistical distributions of various intermediate statistics, Ann. Phys. {\bf 332}, 166-179 (2013).
                  %\bibitem{rev2}
 \bibitem{AlginSenay2012} Algin, A., Senay, M.: High-temperature behavior of a deformed Fermi gas obeying interpolating statistics, Phys. Rev. E {\bf 85}, 041123-(1-10) (2012).
 \bibitem{AdGa} Adamska, L.V., Gavrilik, A.M.: Multi-particle correlations in $qp$-Bose gas model, J. Phys. A: Math. Gen. {\bf 37} 4787-4796 (2004).
 \bibitem{GaMi2012} Gavrilik, A.M., Mishchenko, Yu.A.: Exact expressions for the intercepts of $r$-particle
 momentum correlation functions in $\mu$-Bose gas model, Phys. Lett. A {\bf 376}, 2484-2489 (2012).
 \bibitem{Arik1992Fib} Arik, M. {\it et al}: Fibonacci oscillators, Z. Phys. C {\bf 55}, 89-95 (1992).
 \bibitem{GKR_Fib} Gavrilik, A.M., Kachurik, I.I., Rebesh, A.P.: Quasi-Fibonacci oscillators,
 J. Phys. A: Math. Theor. {\bf 43} 245204 (1-16) (2010).
 \bibitem{Chung_S-TD} Chung, W.S., Gavrilik, A.M., Kachurik, I.I., Rebesh, A.P.: The symmetric Tamm-Dancoff $q$-oscillator:
 the representation, quasi-Fibonacci nature, accidental degeneracy and coherent states, J. Phys. A: Math. Theor. {\bf 47}, 305304 (pp.1-14) (2014).
 \bibitem{OdakaTD1} Odaka, K., Kishi, T., Kamefuchi, S.: On quantization of simple harmonic oscillators,
 J. Phys. A: Math. Gen. {\bf 24}, L591-L596 (1991).
 \bibitem{Chaturvedi1993_TD2} Chaturvedi, S., Srinivasan, V. and Jagannathan, R.: Tamm-Dancoff
 deformation of bosonic oscillator algebras, Mod. Phys. Lett. A {\bf 8}, 3727-3734 (1993).
 \bibitem{GR_TD3} Gavrilik, A.M., Rebesh, A.P.:  A $q$-oscillator with ``accidental'' degeneracy of
 energy levels, Mod. Phys. Lett. A {\bf 22}, 949-960 (2007).
 \bibitem{Bied} Biedenharn L.C.: The quantum group $SU_q$(2) and a $q$-analog of the boson
 operators, J. Phys. A: Math. Gen. {\bf 22}, L873-L878 (1989).
\bibitem{Macfar} Macfarlane A.J.: On $q$-analogues of the quantum harmonic oscillator and
 the quantum group $SU_q$(2), J. Phys. A: Math. Gen. {\bf 22}, 4581-4585 (1989).
 \bibitem{ChapmanHeinz1994} Chapman, S., Heinz, U.: HBT correlators - current formalism vs. Wigner
 function formulation, Phys. Lett. B {\bf 340}, 250-253 (1994).
 \bibitem{KacCheung} Kac, V., Cheung, P.: Quantum Calculus, Springer, Berlin (2002).

\end{thebibliography}
\end{document}